\documentclass{aa}
\usepackage{graphicx}

\begin{document}

\title{Dissipationless Collapse of Spherical \\
       Protogalaxies and the Fundamental Plane}

\author{Christine C. Dantas\inst{1,4}
\
\and Hugo V. Capelato\inst{1}
\and Reinaldo R. de Carvalho\inst{2}
\thanks{\emph{Present adress:} Departamento de Astronomia,
Instituto Astron\^omico e Geof\'{\i}sico,
Universidade de S\~ao Paulo, 01060-970, SP, Brazil.}
\and Andr\'e L. B. Ribeiro\inst{3}}

\institute{Divis\~ao de Astrof\'{\i}sica, INPE/MCT, CP 515, S. J.
dos Campos, SP 12201-970, Brazil.
\and Observat\'orio Nacional, Rua Gal.  Jos\'e Cristino,
77 -- 20921-400, Rio de Janeiro, RJ., Brazil.
\and Departamento de Matem\'atica
Aplicada, IMECC, Universidade Estadual de Campinas,
13083-970 SP, Brazil.
\and Departamento de Astronomia,
Instituto Astron\^omico e Geof\'{\i}sico,
Universidade de S\~ao Paulo, 01060-970, SP, Brazil.}

\date{Received 28 June 2001 / Accepted 4 December 2001}

\maketitle
 
\abstract{
Following on from the numerical work of Capelato, de Carvalho \&
Carlberg (1995, 1997), where dissipationless merger simulations
were shown to reproduce the ``Fundamental Plane" (FP) of elliptical
galaxies, we investigate whether the end products of pure, spherically
symmetric, one-component dissipationless {\it collapses} could also
reproduce the FP.  Past numerical work on collisionless collapses have
addressed important issues on the dynamical/structural characteristics
of collapsed equilibrium systems. However, the study of collisionless
collapse in the context of the nature of the FP has not been
satisfactorily addressed yet. Our aim in this paper is to focus
our attention on the resulting collapse of simple one-component
spherical models with a range of different initial virial
coefficients.  We find that the characteristic correlations of the
models are compatible with virialized, centrally homologous systems.
Our results strengthen the idea that merging may be a fundamental
ingredient in forming non-homologous objects.
}

\section{Introduction}

Self-gravitating stellar systems, ranging from globular clusters to
clusters of galaxies, show significant correlations among kinematic and
photometric parameters (\cite {bur97}).  These correlations are
important tracers of the formation histories of these structures.
However, the origin of these scale relations are not yet known
clearly. 

For instance, the ``Fundamental Plane'' (FP, c.f. \cite{djo87},
\cite{dre87}) relation
of elliptical galaxies represent a significant departure from the
prediction of the virial theorem, under the assumption that ellipticals
are simple, one-component, homologous systems.  This relation is
described by: $r_e \sim \sigma_0^A I^B$ (where $\sigma_0$ is the central
velocity dispersion, $I$, the average surface brightness within the
effective radius in linear units, and $r_e$ is the effective radius,
where $A \sim 1.53$, $B \sim -0.79$ (e.g. \cite{pah98}).  One current
hypothesis to explain the observed discrepancy postulates that the
mass-luminosity ratio of ellipticals would be a function of total
luminosity (e.g. \cite{djo93}, \cite{djo88}, \cite{pah95}). An
alternative hypothesis takes into consideration the possible effects of
the dark matter halo on the FP correlations (\cite{dan00}). Another
explanation (e.g., \cite{hjo95}) is based on the assumption that the
homology hypothesis is not valid, so that elliptical galaxies would be
non-homologous virialized systems.  Several works have addressed the
latter possibility (e.g., \cite{cio96}, \cite{bus97}, \cite{gra97},
\cite{bek98}).  In particular Capelato, de Carvalho \& Carlberg
(1995, 1997, hereon CdCC95 and CdCC97) showed that the FP correlations
arise naturally from objects that are formed by dissipationless
hierarchical mergers of pre-existing galaxies.  The end product of
their simulations was a non-homologous family of objects
following almost exactly the observed K-band FP, with a scatter that
was only half of that observed.

Merging is a natural process in a hierarchical galaxy formation scenario,
and the observed structural properties of ellipticals seem to be well
accounted for by this mechanism (e.g. \cite{shi98}, \cite{ben99} and
references therein).  On the other hand, theoretical and numerical
investigations on pure dissipationless collapses of stellar systems
have historically been of special interest (\cite{pol81}, \cite{hen73},
\cite{alb82}, \cite{mcg84}, \cite{vil84}, \cite{may84}, \cite{mer85},
\cite{agu90}, \cite{lon91}, \cite{kat91}, \cite{car95}).  
In particular, the role of gravitational
instabilities in a model evolving towards equilibrium is still an open
field of investigation (\cite{pal94}), and and may be a key feature for understanding the basics of the dynamics that play a
role in the formation of galaxies. 

However, previous numerical works have not fully addressed the
dynamical/structural characteristics of collapsed systems in the
context of the FP. It is also a question for investigation whether the
results of the merger simulations by CdCC95 can be reproduced
by other initial conditions, that is, whether the FP could arise
solely as a function of the dynamics of relaxation. A
preliminary discussion of these subjects has been given in CdCC97 where
it is suggested that, under appropriate initial conditions,
dissipationless collapses could also follow a FP relationship. We
believe that a first move towards answering the questions raised
above should focus on the simplest dynamical conditions. For this
reason, we restrict our analysis to the classical scenario of
elliptical galaxy formation through one-component dissipationless
gravitational collapse.  Evidently, the effects of a second component
(extended dark halo) on the final equilibrium conditions of the
luminous matter, as much as the presence of gas dissipation, should be
considered in a more refined analysis.

This paper is organized as follows: in Section 2, we
present the simulations and define the characteristic parameters;
in Section 3, the end products of the simulations are considered in
the context of the FP analysis, along with their dynamical/structural
properties. In section 4, we discuss our results.

\section{Simulations Setup and Definition of Characteristic Parameters}

The collapse simulations were run using three different sets of initial
models, denoted by K, A and C. Each set gives a basic
phase-space configuration from which we constructed a grid of collapse
``progenitors'', that is, of out-of-equilibrium models serving as
initial conditions for collapse simulations. This grid was obtained
by pertubing the velocity distributions of the basic set of models, by
re-scaling the particle speeds by constant factors $\sqrt \beta$, where
$\beta \equiv 2T_0/|W_0|$ is the virial ratio of the progenitor, that
is, its ``collapse parameter'' ($T_0$ and $W_0$ are,
respectively, the initial kinetic and potential energies of the
models). The units were chosen as in CdCC95: G = 1, $m_{unit} =
10^{10}$ $M_{\odot}$ and $l_{unit} = 1\rm kpc$, implying $V_{unit} =
207.2 \rm km s^{-1}$ and $T_{unit} = 4.72 \rm Myr$). All the models
have same total mass $M = 20$. The three basic sets were defined as
follows:

\begin{itemize}
\item

``K'' MODELS: Constructed from 8192 Monte Carlo realizations of an equilibrium
spherical King model generated with central potential $W_0 =
5 $ and half-mass radius $r_h = 4$ (see \cite{kin66}). The total radius
of these models is $R = 21.4$. 

\item
``A'' MODELS: Constructed from spherical $\rho \propto r^{-1}$ models of
16384 Monte Carlo particle realizations. Velocities were attributed to
the particles according to an isotropic normal distribution with
velocity dispersion such that $\beta = 1$. The total radius is $R =
20.0$, giving an half-mass radius $r_h$ = 14.1. The initial conditions
obtained from this model are similar to those of the collapse
simulations discussed by Aguilar \& Merritt (1990).

\item
``C'' MODELS: Constructed according to \cite{car95}, with 4096
particles. These are also spherically symmetric models of total radius
$R = 100$, with uniform particle distribuiton $\bar{\rho}$, locally
disturbed by fluctuations with a power spectrum obeying the law
$|\delta_k|^2 \equiv \left| \left({{\rho - \bar{\rho}} \over
\bar{\rho}} \right) \right|^2 \propto k^{n}$, where we used the values
of $n = 1$ (Harisson-Zel'dolvich spectrum; C01-C10 models), $n=0$
(white noise spectrum; C11-C20 models), and $n=2$ (particles-in-boxes
spectrum; C21-C30 models). The peculiar velocities of particles
were distributed as for Models A above and a pure Hubble flow was added
(we adopted $H_0 = 65$ km s$^{-1}$ Mpc$^{-1}$).

\end{itemize}

The grid of collapse progenitors was generated from $0 ~{~}^{<}_{\sim}~
\beta ~{~}^{<}_{\sim}~ 1$. Herein, we generically denote
``cold'' collapses as those resulting from $\beta \rightarrow 0$
progenitors, and ``hot'' collapses those from $\beta \rightarrow 1$.
The complete set of collapse simulations is given in Table
\ref{tab_colapsos}.

\begin{table*}[htbp]
\centering
\caption{Collapses (one-component, DCdCR01)} \label{tab_colapsos}
\vspace {0.3cm}
\begin{tabular}{cccccccccccccc} \hline
\multicolumn{2}{c}{K Models} & \multicolumn{1}{c}{} & \multicolumn{2}{c}{A Models} & \multicolumn{1}{c}{} & \multicolumn{8}{c}{C Models} \\
\cline{1-2} \cline{4-5} \cline{7-14} \\
\multicolumn{2}{c}{$N_{part} = 8192$} & \multicolumn{1}{c}{} & \multicolumn{2}{c}{$N_{part} = 16384$} & \multicolumn{1}{c}{} & \multicolumn{8}{c}{$N_{part} = 4096$} \\
\multicolumn{2}{c}{} & \multicolumn{1}{c}{} & \multicolumn{2}{c}{} &  \multicolumn{1}{c}{} & \multicolumn{2}{c}{$n=1$} & \multicolumn{1}{c}{} & \multicolumn{2}{c}{$n=0$} & \multicolumn{1}{c}{} & \multicolumn{2}{c}{$n=2$} \\ \hline \hline
$\log \beta$ & Run & & $\log \beta$ & Run & & $\log \beta$ & Run & & $\log \beta$ & Run & & $\log \beta$ & Run  \\ \hline
-4.00 & K01  & & -4.00  &  A01   & & -3.75 & C01  & & -3.75 & C11  &  & -3.75 & C21   \\
-3.75 & K02  & & -3.50  &  A02   & & -3.50 & C02  & & -3.50 & C12  &  & -3.50 & C22   \\
-3.50 & K03  & & -3.00  &  A03   & & -3.25 & C03  & & -3.25 & C13  &  & 3.25  & C23   \\
-3.25 & K04  & & -2.50  &  A04   & & -3.00 & C04  & & -3.00 & C14  &  & -3.00 & C24   \\
-3.00 & K05  & & -2.00  &  A05   & & -2.50 & C05  & & -2.50 & C15  &  & -2.50 & C25   \\
-2.75 & K06  & & -1.50  &  A06   & & -2.00 & C06  & & -2.00 & C16  &  & -2.00 & C26   \\
-2.50 & K07  & & -1.25  &  A07   & & -1.50 & C07  & & -1.50 & C17  &  & -1.50 & C27   \\
-2.25 & K08  & & -1.00  &  A08   & & -1.00 & C08  & & -1.00 & C18  &  & -1.00 & C28   \\
-2.00 & K09  & & -0.75  &  A09   & & -0.90 & C09  & & -0.90 & C19  &  & -0.90 & C29   \\
-1.75 & K10  & & -0.50  &  A10   & & -0.80 & C10  & & -0.80 & C20  &  & -0.80 & C30   \\
-1.50 & K11  & & -4.10  &  A01b  & & -4.00 & C01b & &       &      &  &       &           \\
-1.25 & K12  & & -3.60  &  A02b  & & -3.60 & C02b & &       &      &  &       &             \\
-1.00 & K13  & & -3.40  &  A03b  & & -3.40 & C03b & &       &      &  &       &            \\
-0.75 & K14  & & -3.10  &  A04b  & & -3.10 & C04b & &       &      &  &       &            \\
-0.50 & K15  & & -2.75  &  A05b  & & -2.25 & C06b & &       &      &  &       &            \\
-0.25 & K16  & & -2.25  &  A06b  & & -1.75 & C07b & &       &      &  &       &            \\
-0.01 & K17  & & -1.75  &  A07b  & & -1.25 & C08b & &       &      &  &       &          \\ 
      &      & & -1.25  &  A08b  & & -0.25 & C09b & &       &      &  &       &          \\ 
      &      & & -0.95  &  A09b  & & -0.10 & C10b & &       &      &  &       &          \\ 
      &      & & -0.85  &  A10b  & &       &      & &       &      &  &       &          \\ 
      &      & & -0.25  &  A11   & &       &      & &       &      &  &       &          \\ 
      &      & & -0.10  &  A12   & &       &      & &       &      &  &       &          \\

\hline
\end{tabular}
\end{table*}

The simulations were run using a C translation of the \cite{bar86}
TREECODE, running on {\it Ultra Sparc} and  {\it Sparc 5} machines.
Quadrupole correction terms, according to \cite{dub88}, were used in
the force calculations. We set the tolerance parameter, time step and
potential softening length as 0.8, 0.025 and 0.05 respectively, as
described in CdCC95. In particular, the softening length ($\epsilon$)
was carefully chosen in order to conform to the constraints of resolution
and collisionlessness, given the total number of particles
($N_{part}$) used in each simulation.  According to Barnes \& Hut
(1989), structural details up to scales $\sim 10\epsilon$ are sensitive
to the value of $\epsilon$. Thus the spatial resolution of a simulation
must satisfy $\lambda > 10\epsilon$.  We set the resolution of our
simulations to $\lambda < 0.2 r_e$, where $r_e$ is the effective radius
of the simulated models (see below).  This results in the constraint
$\epsilon < r_e/50$. On the other hand, in order to achieve the
collisionless condition, $\epsilon$ must exceed, by a certain factor,
the typical scale where important collisions occur, that is $\epsilon >
C p_{90}$, where $p_{90}$ is the impact parameter for a $90^{\rm o}$
deviation in encounters between two particles in hyperbolic orbit,
$p_{90} \sim Gm/{\bar{v}}^{2}$, where $m$ is the mass of particles in
the simulation and $\bar{v}$ is the mean velocity dispersion of the
simulated model.  C is a heuristic factor found to be in the range
50-100 (see discussion in \cite{bar89}).  Notice that these
criteria also imply $C p_{90} < \lambda/10$, allowing lower bounds for
the number of particles $N_{part}$. Upper bounds are set by the total
CPU times which, for the tree-code used here, varies as $ \sim
N_{part}\log N_{part}$.  An increase in the number of particles is
always interesting because it increases the spatial resolution of the
simulation. Thus, in the case of measuring central velocity
dispersions, better statistics are expected if more
particles are introduced. In any case, the {\it general} results should
be reasonably unchanged by the use of different numbers of particles
(for instance, by using $N_{part} = 5000$ instead of a number $10$ times greater). 
In our work, we carefully adjusted both parameters, $\epsilon$ and
$N_{part}$ in order to conform to both the spatial resolution and
collisionless constraints as well the operational constraints due to
CPU times. Among the simulations summarized above, the C models were
the most CPU time-consuming, as they include the evolution since the
initial decelerated expansion phases, before the turn-around and
collapse of the system. This forced us to use smaller values of
$N_{part}$ in this case. These lower values, however, are well above
the lower bound discussed before.

The models evolved up to $\sim 30$ ``crossing times''
($T_{cr} = GM^{5/2}/(2E)^{3/2}$), when quantities like half-mass radius
and the virial ratio, $\beta$, indicated no significant variation in
the system.
We point out that, although the initial conditions of the
C models are of cosmological relevance, the evolution of
these models do not represent cosmological simulations {\it per se}
(i.e., the evolution of small density pertubations in a large
volume of the primordial universe). The aim of these models is
to represent a spherically symmetric fluctuation that detached from
the general expansion. In this sphere, we assume that the stellar
formation occurred over a short time scale, so that, as with the
other models, we are representing a system that will evolve through
pure stellar dynamics. 

\section{The Fundamental Plane of End Products}

\begin{figure*}
\centering
\includegraphics[width=12cm]{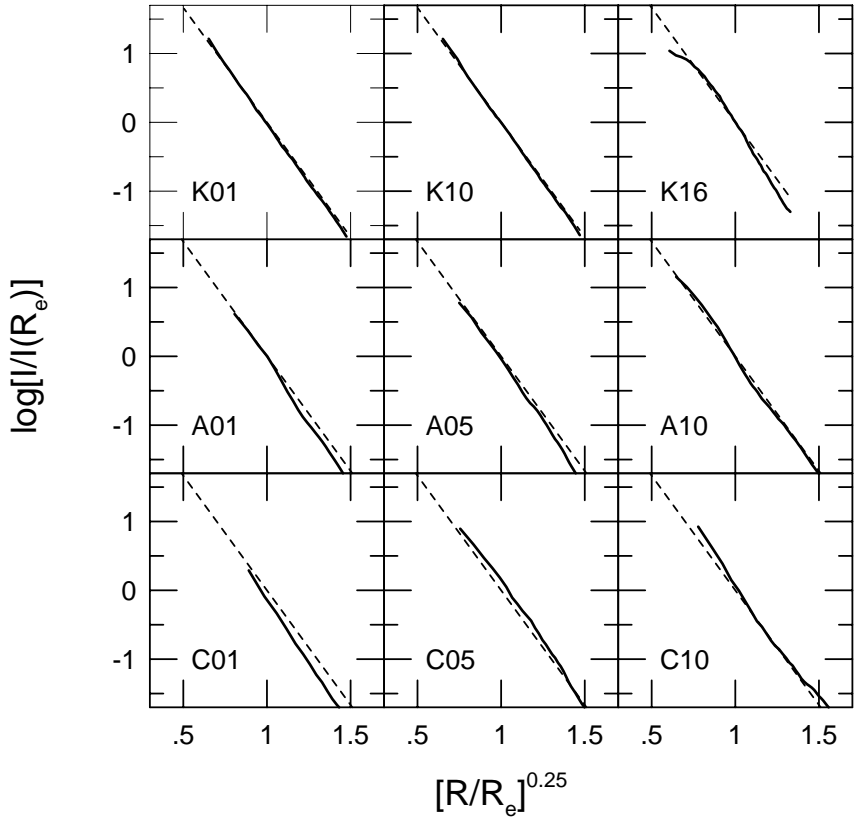}
\caption{Surface density profiles (solid lines) 
of some examples of collapses, in comparison to the
de Vaucouleurs law (dotted lines).}
\label{prof_col}
\end{figure*}

Before analysing the equilibrium-simulated objects in the
context of the FP relations, we obtained the 
surface density profiles of the final models and
qualitatively compared them to the 
de Vaucouleurs (1948) profile. The objects typically
show a reasonable concordance with this profile. We present
some examples in Fig.\ref{prof_col}.

In order to reproduce the observed quantities related to the FP, we
followed the procedure given by CdCC95 to compute the characteristic
FP variables of the simulated models:  the effective radius $r_e$,
containing half of the total projected mass of the system; the mean
surface density within $r_e$, $\Sigma_e = {M(<r_e)/\pi r_e^2}$, so that
$\mu_e \equiv -2.5 \log \Sigma_e$ and the central projected velocity
dispersion, $\sigma_0$.  These quantities were estimated as the
{\sl median} of 500 randoms projections of the collapsed objects so that the
associated error bars reflect the expected dispersions due to
projection effects.  $\sigma_0$ and $\mu_e$ where combined in the
vertical axis according to the usual representation of the FP projected
onto a Cartesian plane with $r_e$ on the horizontal axis (c.f.
CdCC95).

\begin{figure*}
\centering
\includegraphics[width=12cm]{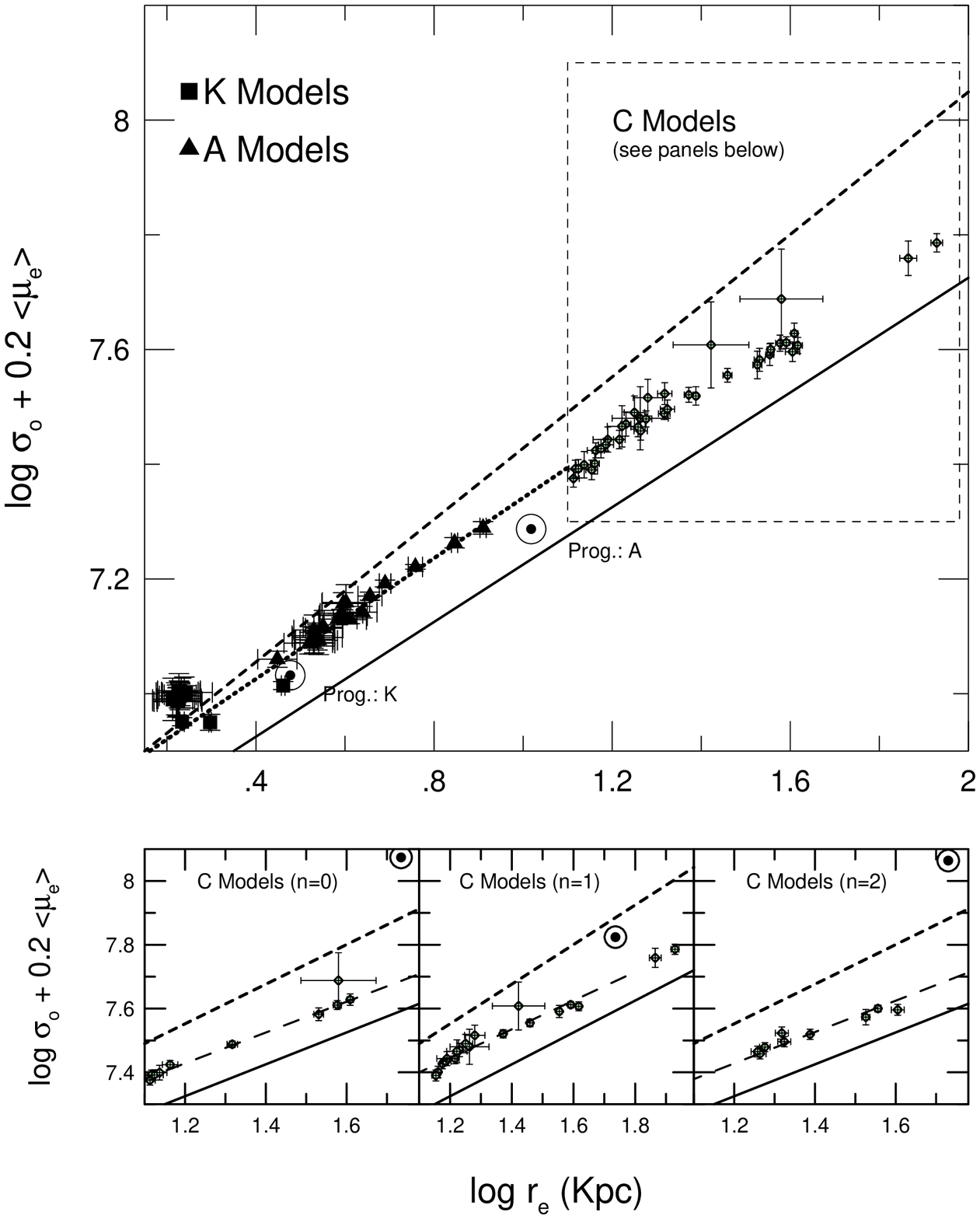}
\caption{Results of the collapse simulations in terms of the FP parameters.  The
positions of the equilibrium models corresponding to the three basic
sets of initial models are shown circumscribed by larger circles. The
solid line gives the slope a family of homologous equilibrium
systems would have in this plot, whereas the short dashed line is the best fit
to the mergers products of CdCC95. For clarity 
the C models have been plotted separately on the three lower
panels.  {\it Main (top) figure:} The dotted line fits the A models.
The C models are shown for direct comparison with the K and A models.
The dashed square represents the area traced by the smaller panels
below.  {\it Lower panels:} Detailed results for the C models, where
each panel shows the C models in terms of their $n$ parameter. The long
dashed line fits each of the C models.}
\label{fp}
\end{figure*}

Figure \ref{fp} shows the results of the collapse simulations in terms
of the FP parameters. The positions of the ``$\beta = 1$
progenitors'', that is, of the equilibrium models of the three basic
sets of initial models, are circumscribed by larger circles. The
positions of the collapse progenitors lay in the $\log r_e = cte$ lines
passing by these points, at distances of about $log \sigma_o
\sim log \sqrt \beta$ downwards. In Table \ref{tab_ajus_col},
we present the best fit values of the resulting slopes ($\alpha$) of
the scaling relations for each family of models (the K models were not
adjusted for the reasons explained below).  

We also present in this Table the best fit results for the
``coldest'' ($\log \beta \leq -2.00$) C n=1 collapses as well. The
reason for this decomposition is based on a visual inspection from
Figure \ref{fp} (lower, middle panel), which suggests that there may
exist an underlying difference for the ``coldest'' C n=1 collapses, in
the sense that these two groups might be separately non-homologous,
although they globally (viz. in conjunction) fit a relation close to
homology.

\begin{table*}[htbp]
\centering
\caption{FP ``Tilt'': Best Fits} \label{tab_ajus_col}
\vspace {0.3cm}
\normalsize
\begin{tabular}{lcc} \hline
\multicolumn{1}{c} {MODEL} & \multicolumn{1}{c} {$\alpha \pm \delta\alpha$} & \multicolumn{1}{c} {$N_{fit}$}  \\ 
\hline \hline 
K Collapses                                   & -                          & -   \\ \hline
A Collapses                                   & $\alpha = 1.954 \pm 0.123$ & 22  \\ \hline
C Collapses:                                  &                            &     \\
n = 0                                         & $\alpha = 2.070 \pm 0.123$ & 10  \\
n = 1 (all collapses)                         & $\alpha = 2.161 \pm 0.087$ & 19  \\ 
n = 1 (only $\log \beta \leq -2.00$ models)   & $\alpha = 1.070 \pm 0.270$ & 11  \\ 
n = 2                                         & $\alpha = 2.033 \pm 0.342$ & 10  \\ \hline
\end{tabular}
\end{table*}

First, by examining this figure and the best fitting slopes given in
Table \ref{tab_ajus_col} we find that the FP correlations (illustrated
by the short dashed line in the figure) do {\it not} seem to
be recovered by the collapsed systems. In fact, 
the best fitting slopes are all consistent
with $\alpha =2$ which is the slope predicted by the virial equations
applied to homologous systems (represented by the solid line in the
figure).  Second, the collapsed objects tend to cluster in the FP space
variables with their location depending primarily on the initial model
from which they result.

Indeed, the K models cluster in a small region of the FP space, except
for two or three ``hottest'' collapses ($\log \beta > -0.75$), which
detach from the ``cluster''.  This somewhat precludes any attempt to
fit the resulting objects into a scaling relation. The A
collapses also show this behaviour, but not as strongly as the K
models. In the case of A collapses,  the best fit using all the points
(dotted line in the figure) gives a relation close to that expected
for homologous systems ($\alpha = 1.900$).

The clustered distribution of the K models is a peculiar feature
among the general behaviour of the collapsed models in the FP parameter
space. The K models do not ``spread'' along the virial plane, whereas
the other models are well distributed.  A naive approach to understand
the reasons for this different behaviour is the following: by a
combination of the energy conservation and the virial theorem, an
arbitrary distribution of stars with initially zero random velocities
will collapse and settle in a configuration with a radius typically a
half of the initial radius.  Evidently, our models do not start with
such extremely ``cold'' conditions, since the progenitors are in fact
assigned with different values of their collapse factors (the $\beta$
parameter, as described in Section 2). This should spread the models
along the virial surface.  However, with the abovementioned
approximation in mind, it is expected that the K and A models, which
have the same initial radius ($R=20$), should occupy after
virialization similar locations in $\log r_e$, whereas the C models
(initial radius $R=100$) should settle in larger $r_e$'s.  In fact, the
C models are located in the upper right region of the diagram in Fig.
\ref{fp}. The A and K models, on the other hand, do not occupy similar
locations and distributions in the FP parameter space, although they
are assigned the same collapse factors as the other models.  The K and
A models have different initial mass distributions, as can be seen by
their different initial effective radii.  It is possible that the
initial central mass concentration of a given model may dynamically
influence the final locations of the models within the FP space, as
well as driving the models towards a more clustered distribution on the
virial plane. Unfortunately, a more descriptive argument on how the
dynamics could play that role is not clear at present.

The C models, quite independently from the spectral index $n$, may
globally approach homologous systems.   By keeping 
only the ``coldest'' C n=1 models, the resulting fit 
deviates from homology ($\alpha = 1.070$), although 
the associated error bar is large ($N_{fit} = 11$).
In any case, these results substantially differ from those found for
the set of merger simulations of CdCC95, where the FP relation is fully
recovered, as can be seen by the short dashed line on Figure \ref{fp},
which represents the best fit for the merger end products (slope
$\alpha = 1.36 \pm 0.08$), as given in their paper.  

One word of caution concerns the somewhat large error bars
resulting from some of the fits to our collapse models in the FP
parameter space, compared to that found by CdCC95 ($\delta \alpha =
0.08$; the slope error quoted in CdCC95 results from a fit to $17$
models).  We have confirmed that this effect results from the number of
models used for the corresponding fits, which are, in some cases, small
(e.g., only $10$ runs for the n=0, n=2 C models). We have increased the
number of runs for the A (initially from $10$ to $22$ runs) and for the
C n=1 models (from $10$ to $19$ runs) and the error bars ($\delta
\alpha$'s) decreased by $\sim 40-50 \%$.  The general result 
is that the final collapsed models do not seem to follow a
FP-like relation.

In order to further assertain the general homologous nature of the
models, we follow the analysis of CdCC95.  The virial
theorem, applied to a stationary, self-gravitating system, establishes
that $\langle v^2\rangle = {GM/r_G}$, where $r_G$ is the gravitational
radius (c.f. \cite{bin87}) and $\langle v^2 \rangle^{1 \over 2}$ the
three-dimensional velocity dispersion.  These physical quantities can
be translated into observational ones through a definition of certain
kinematical-structural coefficients ($C_r, C_v$), which may or may not
be constants among galaxies:

\begin{equation}
C_r \equiv r_G/r_e
\end{equation}
and
\begin{equation}
C_v \equiv \langle v^2 \rangle / \sigma_0^2;
\end{equation}
Defining $I_e \equiv \eta \Sigma_e$, with $\eta \equiv \left ( {M \over
L} \right )^{-1}$, and inserting the equations above into the
virial relation, we find that $r_e = C_{vir} \eta \sigma_0^2 I_e^{-1}$,
where:
\begin{equation}
C_{vir} \equiv {C_r C_v \over 2 \pi G }.
\end{equation}

\begin{figure*}
\centering
\includegraphics[width=12cm]{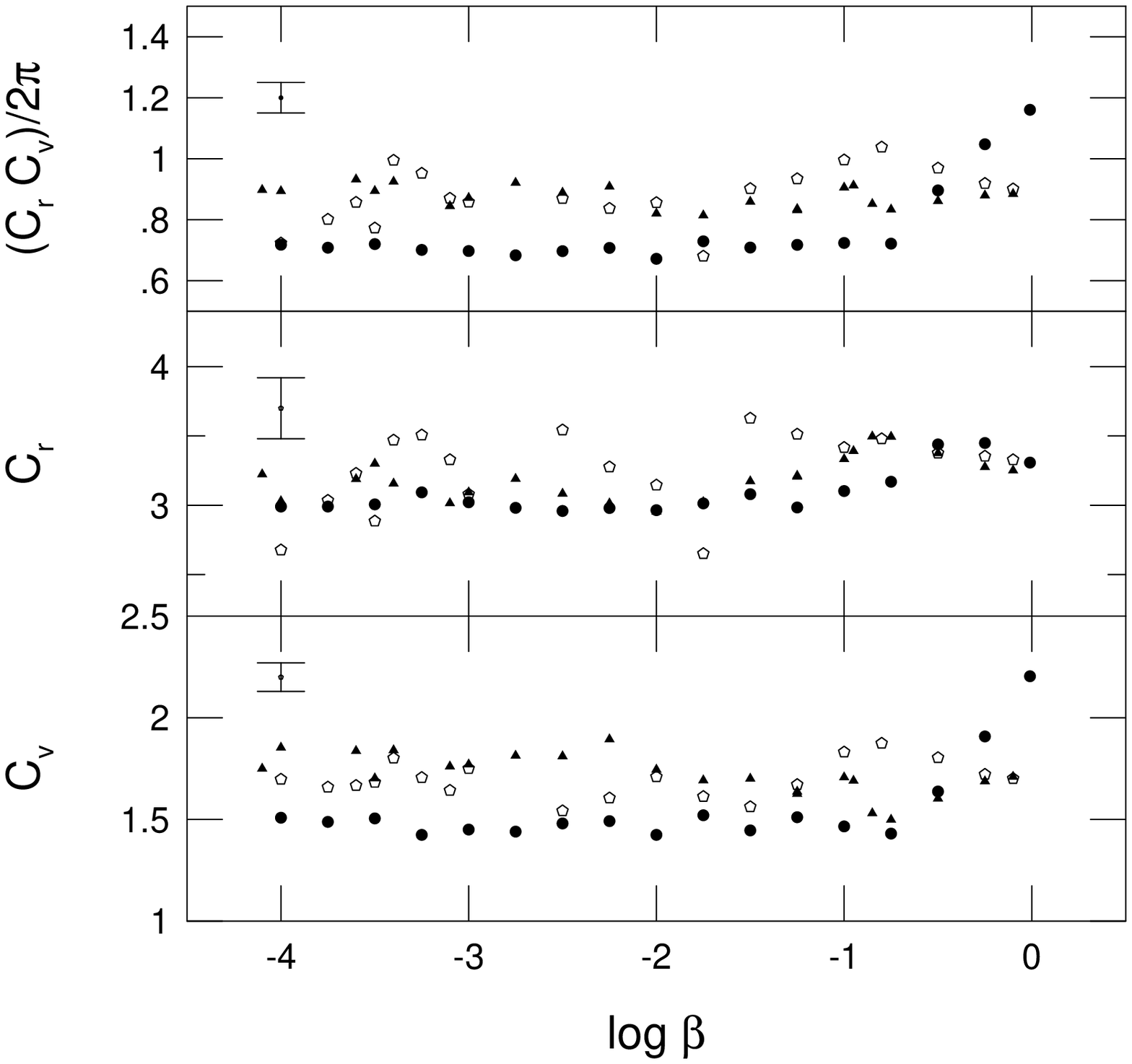}
\caption{Ratio of the virial coefficients
$C_{vir}$ (upper panel), $C_r$ (central panel)
and $C_v$ (lower panel)
as a function of the initial condition
$\log \beta$. Symbols: solid circles for the K collapses,
solid triangles for the A models, and open pentagons for the
C n=1 models.}
\label{homol}
\end{figure*}

In the case of our numerical simulations, $\eta$ should be taken as a
constant. We computed the final structural/kinematical virial coefficients defined above and the results are
presented in Figure \ref{homol}, where we plot these virial coefficients
as a function of the collapse parameter.  The associated
errors are represented by the vertical bars and the symbols are the
same of Figure \ref{fp}. For the C models, we limit ourselves
to the $n=1$ case, which represents the strongest deviation
from homology for the ``coldest'' initial conditions.

We find that, in fact, the collapsed K and A models are approximately
homologous among themselves, since the structural/kinematical virial
coefficients do not vary considerably among them.  Any residual
non-homology present is due to the presence of the ``hottest''
collapses. On the other hand, the C ($n=1$) models may be
thought of as globally homologous objects, but the virial coefficients
do show some significant noise, considering the ``coldest'' family of
collapses. This is compatible with what has previously been deduced
from the FP parameter space.

\section{Discussion}

We designed a set of numerical experiments of dissipationless collapse
of spherical models in order to investigate some possible constraints
on the origin of the FP of elliptical galaxies.  

First, the final objects are found at different locations on the FP
parameter space, depending on the initial model. Even if the initial
models have the same mass and length (e.g. $R$) scales, which is the case
for the K and A models, they will occupy almost non-intersecting regions
on the FP space. The central mass concentration of the initial models
may have a determining influence on the final locations of the models on
the FP space, for the same set of global structural-dynamical parameters
(viz. $\beta$ and $R$).

Second, we found that the collapsed models do not globally populate the
FP, being more compatible with homologous virialized systems.  In
particular, the final ``coldest'' C models, especially the n=1 models,
tend to form a family with a deviation from homology, although the
error bars are too large to confirm this effect. The C models are not
only the result of a global collapse of the system, but also the
product of a series of mergers of small fluctuations of smaller scales,
which had the opportunity to grow depending on the initial conditions.
In particular, the $n=0$ models seem to have collapsed more
homogeneously, since its resulting scaling relations slope is quite
similar to the K and A models, which are initially spherically
symmetric objects with diferent central mass concentrations. In fact,
the $n=0$ models represent a homogeneous model where small-scale
noise has been added, so that sufficiently large clumps probably were
not formed. This is unlike the $n=1$ and $n=2$ models, that might have
resulted from mergers of significant clumps along with the main global
collapse.

From the above discussion and also from the results of CdCC95, we
conclude that merging may indeed be an important ingredient in
forming non-homologous objects. In addition, 
we consider that, although ``hot'' collapses result in a
``softer'' process toward virialization (so that the relaxation
mechanism of these collapses should be more closely related to the one
produced by a slow merger), it is the mechanism of merging {\it per se}
that drives objects towards non-homology. This is because we
 found that ``cold'' collapses from initial clumpy conditions are
non-homologous, whereas the same cold collapse factors applied to more
spherically symmetric systems {\it do} produce homology. From this, we
also deduce that any dynamical constraints dictated by spherical
symmetry in the initial conditions are not likely to have 
been present
in the formation of elliptical galaxies, at least for those
in clusters.  This result is of course
consistent with those found in the classical literature, where
reasonably strong indications are found that ``clumpy''
and ``cold'' initial conditions are fundamental for the formation of
objects similar to ellipticals (e.g.  \cite{alb82}, \cite{may84},
\cite{mer85}, \cite{agu90}, \cite{lon91}).

\begin{figure*}
\centering
\includegraphics[width=12cm]{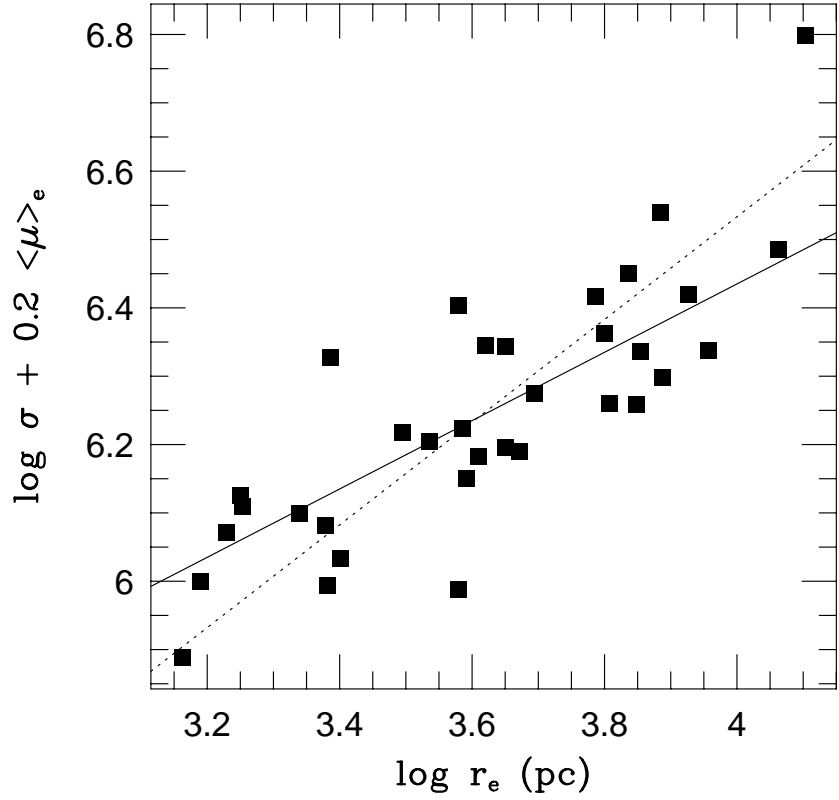}
\caption{Field ellipticals data from
de Carvalho \& Djorgovski (1992) in terms of the FP parameters. 
The solid line is a homologous fit and the dotted line is a FP fit 
for the Coma cluster.}
\label{elipcampo}
\end{figure*}

On the other hand, considering 
the possibility that {\it field} elliptical galaxies
underwent insignificant merging episodes in
their past histories (e.g. \cite{tot98}), and
assuming that their gross properties could be 
described by a simple, spherically symmetric collapse model, our simulations
indicate that they should present a FP relation much closer to the
virial expectations, compared to cluster ellipticals.
Alternatively, these field elliptical galaxies could also 
be non-homologous objects if they collapsed from 
reasonably ``clumpy'' and ``cold'' initial conditions.
Recently, some authors have concentrated on
the analysis of the FP of field early-type galaxies, but the data is
still very preliminary and precludes any firm determination of the
resulting FP ``tilt'' for these galaxies (e.g.,\cite{dok01},
\cite{tre99}, \cite{pah98}, \cite{dcd92}).  For instance, we directly
tested the above hypothesis using the field ellipticals data from
de Carvalho \& Djorgovski (1992). 
In Figure \ref{elipcampo}, we show this data in terms of
the FP parameters. The solid line is a homologous fit  and the dotted
line is a FP fit for the Coma cluster.  The residuals show that it is
not possible to verify which of the relations is the best fit due to
the high dispersion of the data.  The K models in our simulations are
able to produce some dispersion, depending on the initial parameters
(e.g. $\beta$ and/or mass), but they are not able to ``move'' the
resulting models along any relation.  We conclude that much more data
is needed in order to decisively test the constraints presented by our
models on the FP of field ellipticals.

Finally, we point out that our present analysis is, of course, not
entirely realistic, and is based on simple one-component
dissipationless models.  Our conclusions should be interpreted more
as a {\it general trend} to be further investigated by a 
larger series of higher resolution simulations.
But the exercise seems to be useful in
bringing  out some clues that otherwise could be difficult to unravel
in a more complex scenario.  We intend to further refine our results
adopting more realistic scenarios. Two-component models and full
cosmological simulations are being currently investigated and will
be the subject of a future paper.

\begin{acknowledgements}

We thank J. Dubinski for making his ``tree-code'' available for this
project.
We thank the anonymous referee for useful suggestions.
C.C.D. acknowledges fellowships from FAPESP under grants 96/03052-4 and
01/08310-1.  A.L.B.R. acknowledges fellowships from FAPESP under grant
97/13277-6. This work was partially supported by CNPq and FAPESP (Project No.
2000/06695-2).

\end{acknowledgements}


\begin{thebibliography}{}

\bibitem [Aguilar \& Merritt 1990] {agu90} Aguilar, L. A. \& Merritt, D.,
1990, ApJ 354, 33.
\bibitem [Barnes \& Hut (1986)] {bar86} Barnes,J. \& Hut, P.,
1986, Nature v.324, 446.
\bibitem [Barnes \& Hut 1989] {bar89} Barnes,J. \& Hut, P., 
1989, ApJS 70, 389.
\bibitem [Bekki 1998] {bek98} Bekki, K., 1998, ApJ, 496, 713.
\bibitem [Bender \& Saglia 1999] {ben99} Bender \& Saglia, 1999,
in ASP Conf. Ser. 182, Galaxy Dynamics,  
ed. David R. Merritt, Monica Valluri, and J. A. Sellwood.
(San Francisco: ASP), 113.
\bibitem [Binney \& Tremaine 1987]{bin87} Binney, J. \& Tremaine,
S., 1987, Galactic Dynamics (Princeton: Princeton Univ. Press).
\bibitem [Burstein et al. 1997] {bur97} Burstein, D., Bender, R., Faber, S. M.,
\& Nolthenius, R., 1997, AJ, 114, 1365.
\bibitem [Busarello et al. 1997] {bus97} Busarello, G., Capaccioli, M.,
Capozziello, S., Longo, G., \& Puddu, E., 1997, A\&A, 320, 415.
\bibitem [Capelato, de Carvalho \& Carlberg 1995] {cap95}
Capelato, H. V., de Carvalho, R. R. \& Carlberg, R. G., 1995, ApJ,
451, 525.
\bibitem [Capelato, de Carvalho \& Carlberg 1997]{CdCC97} 
Capelato, H.V., de Carvalho, R.R., Carlberg, R.G., 1997, 
in ESO Workshop, Galaxy Scaling Relations: Origins, Evolution and Applications,
ed. L.N. da Costa \& A. Renzini (Springer-Verlag), 331.
\bibitem [Carpintero \& Muzzio (1995)]{car95} Carpintero, D. D. 
\& Muzzio, J. C., 1995, ApJ, 440, 5.
\bibitem [Ciotti, Lanzoni \& Renzini 1996] {cio96} 
Ciotti,L., Lanzoni,B., Renzini, A., MNRAS, 282, 1.
\bibitem [Dantas et al. 2000] {dan00} Dantas, C. C., Ribeiro, A. L. B.,
Capelato, H. V., \& de Carvalho, R. R., 2000, ApJ, 528, L5.
\bibitem [de Carvalho \& Djorgovski 1992]{dcd92} de Carvalho, R.R.
\& Djorgovski, S., 1992, ApJ, 389, L49.
\bibitem [de Vaucouleurs 1948)]{vau48} de Vaucouleurs, G., 1948,
Ann. d'Astrophys., 11, 247.
\bibitem [Djorgovski \& Davis 1987] {djo87} Djorgovski, S. G. \& Davis,
M., ApJ, 1987, 313, 59.
\bibitem [Djorgovski 1988] {djo88} Djorgovski, S. G., 1988, in Proc. Moriond
Astrophysics Workshop, Starbursts and Galaxy Evolution, ed. T. X.
Thuan et al. (Gif sur Yvette: Editions Fronti\`eres), 549.
\bibitem [Djogorvski \& Santiago 1993] {djo93}
Djogorvski, S. G. \& Santiago, B. X., 1993, in 
ESO Conf. and Workshop Proc. 45, Structure, Dynamics and 
Chemical Evolution of Elliptical Galaxies, 
ed. I. J. Danziger, W. W. Zeilinger, \& K. Kjar (Garching:ESO), 59.
\bibitem [Dressler et al. 1987]{dre87} Dressler, A.,
Lynden-Bell, D., Burstein, D., Davies, R. L., Faber, S. M.,
Terlevich, R. J., \& Wegner, G., 1987, ApJ, 313, 42.
\bibitem [Dubinski (1988)]{dub88} Dubinsky, J. 1988, MS thesis, Univ. of
Toronto.
\bibitem [Graham \& Colless 1997] {gra97} Graham, A. \& Colless, M.,
1997, MNRAS, 287, 221.
\bibitem [H\'enon 1973] {hen73} H\'enon, 1973, A\&A 24, 229.
\bibitem [Hjorth \& Madsen 1995] {hjo95} Hjorth, J. \& Madsen, J.,
 1995, ApJ, 445, 55.
\bibitem [Katz 1991] {kat91} Katz, 1991, ApJ 368, 325.
\bibitem [King 1966]{kin66} King, I. R., 1966, AJ, 71, 64.
\bibitem [Londrillo, Messina \& Stiavelli 1991] {lon91}
Londrillo, P., Messina, A., \& Stiavelli, M., 1991, MNRAS, 250, 54.
\bibitem [May \& van Albada 1984] {may84} May \& van Albada, 1984, MNRAS, 209, 15.
\bibitem [McGlynn 1984] {mcg84} McGlynn, 1984, ApJ 281, 12.
\bibitem [Merritt \& Aguilar 1985] {mer85} Merritt \& Aguilar, 1985, MNRAS, 217, 787.
\bibitem [Pahre, Djorgovski \& de Carvalho 1995] {pah95} Pahre, M. A.,
Djorgovski, S. G. \& de Carvalho, R. R., 1995, ApJL, 453, L17.
\bibitem [Pahre, Djorgovski \& de Carvalho 1998]{pah98} Pahre, M. A., Djorgovski, S. G. \& de Carvalho, R. R., 1998, AJ, 116, 1591.
\bibitem [Palmer 1994] {pal94} Palmer, P. L., 1994, Stability of
Collisionless Stellar Systems, Kluwer Academic Publishers.
\bibitem [Polyachenko \& Shukhman 1981] {pol81} Polyachenko \& Shukhman, 1981,
 Sov.Astron. 25, 533.
\bibitem [Shier \& Fischer 1998] {shi98} Shier, L. M. \& Fischer,
J., 1998, ApJ, 497, 163.
\bibitem [Totani \& Yoshii 1998]{tot98} Totani, T. \&
Yoshii, Y., 1998, ApJ, 501, L177.
\bibitem [Treu et al. 1999]{tre99} Treu, T., Stiavelli, M., Casertano, S.,
Moller, P., \& Bertin, G., 1999, MNRAS, 308, 1037.
\bibitem [van Albada 1982] {alb82} van Albada, 1982, MNRAS, 201, 939.
\bibitem [van Dokkum et al. 2001]{dok01} van Dokkum, P. G.,
Franx, M., Kelson, D., \& Illingworth, G. D., 2001, ApJL, in press
(astro-ph/0104155).
\bibitem [Villumsen 1984] {vil84} Villumsen, 1984, ApJ, 284, 75.

\end{thebibliography}
\end{document}